\documentclass[aps,prd,twocolumn,showpacs,superscriptaddress,nofootinbib]{revtex4}

\usepackage{graphicx}
\usepackage{times}
\usepackage[pdfpagemode=UseNone,colorlinks=true,citecolor=black,filecolor=black,linkcolor=black,urlcolor=black]{hyperref}
\usepackage{breakurl}

\begin{document}


\title{Core-Collapse Astrophysics with a Five-Megaton Neutrino Detector}

\author{Matthew D. Kistler}
\affiliation{Department of Physics, Ohio State University, Columbus, Ohio 43210, USA}
\affiliation{Center for Cosmology and  Astro-Particle Physics, Ohio State University, Columbus, Ohio 43210, USA}

\author{Hasan Y{\"u}ksel}
\affiliation{Department of Physics, Ohio State University,  Columbus, Ohio 43210, USA}
\affiliation{Center for Cosmology and Astro-Particle Physics, Ohio State University, Columbus, Ohio 43210, USA}

\author{Shin'ichiro Ando}
\affiliation{California Institute of Technology, Mail Code 350-17, Pasadena, California 91125, USA}

\author{John F. Beacom}
\affiliation{Department of Physics, Ohio State University, Columbus, Ohio 43210, USA}
\affiliation{Center for Cosmology and Astro-Particle Physics, Ohio State University, Columbus, Ohio 43210, USA}
\affiliation{Department of Astronomy, Ohio State University, Columbus, Ohio 43210, USA}

\author{Yoichiro Suzuki}
\affiliation{Kamioka Observatory, Institute for Cosmic Ray Research, University of Tokyo, Hida, Gifu 506-1205, Japan}
\affiliation{Institute for the Physics and Mathematics of the Universe, University of Tokyo, Kashiwa, Chiba 277-8568, Japan}


\date{April 8, 2011}

\begin{abstract}
The legacy of solar neutrinos suggests that large neutrino detectors should be sited underground.  However, to instead go underwater bypasses the need to move mountains, allowing much larger water \v{C}erenkov detectors.  We show that reaching a detector mass scale of $\sim 5$ Megatons, the size of the proposed Deep-TITAND, would permit observations of neutrino ``mini-bursts'' from supernovae in nearby galaxies on a roughly yearly basis, and we develop the immediate qualitative and quantitative consequences.  Importantly, these mini-bursts would be detected over backgrounds without the need for optical evidence of the supernova, guaranteeing the beginning of time-domain MeV neutrino astronomy.  The ability to identify, to the second, every core collapse in the local Universe would allow a continuous ``death watch'' of all stars within $\sim 5$ Mpc, making practical many previously-impossible tasks in probing rare outcomes and refining coordination of multi-wavelength/multi-particle observations and analysis.  These include the abilities to promptly detect otherwise-invisible prompt black hole formation, provide advance warning for supernova shock-breakout searches, define tight time windows for gravitational-wave searches, and identify ``supernova impostors'' by the non-detection of neutrinos.  Observations of many supernovae, even with low numbers of detected neutrinos, will help answer questions about supernovae that cannot be resolved with a single high-statistics event in the Milky Way.
\end{abstract}

\maketitle



\section{Introduction}

Core-collapse supernovae have long been suspected to be the solution of many long-standing puzzles, including the production of neutron stars and black holes, radioactive isotopes and heavy elements, and cosmic rays~\cite{Baade}.  Understanding these issues, and the properties of neutrinos and hypothesized new particles, requires improving our knowledge of supernovae.  It is not enough to record their spectacular visual displays, as these do not reveal the dynamics of the innermost regions of the exploding stars, with their extremes of mass and energy density.  Moreover, sophisticated simulations of the core collapse of massive stars do not robustly lead to supernova explosions~\cite{Buras:2003sn,Burrows:2005dv,Mezzacappa:2005ju}, raising the suspicion that crucial physics is missing.

Neutrinos are the essential probe of these dynamics, as they are the only particle that escapes from the core to the observer (gravitational waves may be emitted, but they are energetically subdominant).  There is an important corollary to this, namely {\it until supernovae besides SN~1987A are detected by neutrinos, our fundamental questions about supernovae will never be decisively answered.}  In fact, the most interesting problems--associated with the presence, nature, variety, and frequency of core collapse in massive stars--can only be solved by detecting {\it many} supernova neutrino bursts.

\begin{figure}[t!]
\includegraphics[width=3.25in,clip=true]{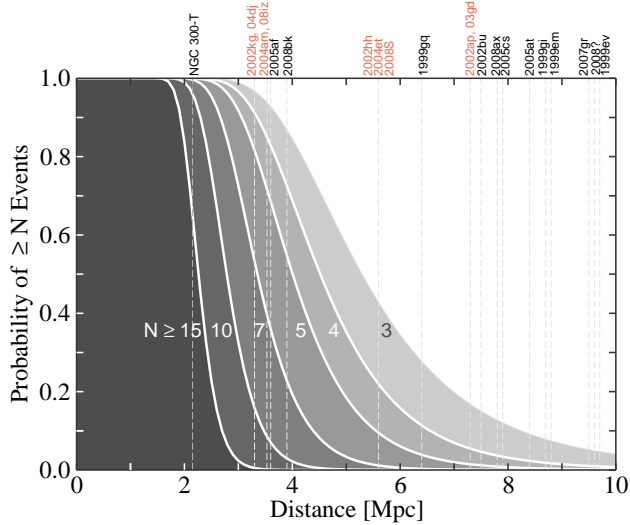}
\caption{Probabilities to obtain the indicated numbers of $\bar{\nu}_e$ neutrino events (with $E_{e^+} > 18$ MeV) in a 5~Mton detector as a function of the supernova distance.  We assume a Fermi-Dirac $\bar{\nu}_e$ spectrum with an average energy of 15 MeV and a total energy of $5\times 10^{52}$~erg.  Optical supernovae observed in the 10 years from 1999-2008 are noted at their distances; those in red indicate multiple supernovae in the same galaxy.}
\label{fig:yields}
\end{figure}

The challenges of supernova neutrino burst detection are that Milky Way sources are rare and that more common distant sources have little flux.  The 32 kton Super-Kamiokande (SK) detector is large enough to detect with high statistics a burst from anywhere in the Milky Way or its dwarf companions, but the expected supernova rate is only 1--3 per century, and there is no remedy but patience.  Proposed underground detectors~\cite{Nakamura:2003hk,Jung:1999jq,deBellefon:2006vq,Autiero:2007zj}, like the $\sim 0.5$~Mton Hyper-Kamiokande (HK), could detect one or two neutrinos from supernovae in some nearby galaxies~\cite{Ando:2005ka}.  As shown in Fig.~\ref{fig:yields}, to robustly detect all neutrino bursts within several Mpc, where recent observations show the supernova rate to be at least $\sim 1$ per year, requires scaling up the detector mass of SK by about two orders of magnitude, to at least $\sim 5$ Mton.

A recent proposal for the Deep-TITAND detector shows in detail how it might be feasible to build such a large detector in a cost-effective way~\cite{Suzuki:2001rb, Suzuki}.  To avoid the high costs and slow pace of excavating caverns underground, this proposal conceives of a modular 5~Mton undersea detector that could be constructed quickly.  Key motivations for such a detector are superior exposure for studies of proton decay, long-baseline neutrinos, and atmospheric neutrinos.  To reduce costs, the detector would be built with a shallower depth and lower photomultiplier coverage than SK; these decisions would sacrifice the low-energy capabilities for all but burst detection.

There is a compelling case for a 5 Mton detector based on supernova neutrino detection alone, and the science benefits that we discuss here will hold even if a Milky Way supernova is detected first.  Individual core collapses could be detected in mini-bursts of neutrino events, with $N \gtrsim 3$ events needed to suppress detector backgrounds.  The expected yields for objects in nearby galaxies are high enough to detect neutrinos when there is no optical display (due to a weak or failed explosion or obscuration) or when the nature of the transient is debated (the so-called ``supernova impostors'').  In addition, the combined data from many bursts would measure the average supernova neutrino emission, which could be compared to the SN~1987A data \cite{Hirata:1987hu,Hirata:1988ad,Bionta:1987qt,Bratton:1988ww} and future data.

Indeed, a 5 Mton supernova neutrino detector is one of the most promising prospects for developing an observatory for non-photon time-domain astrophysics.  The minimal size of the required detector is known now, and it is not out of reach, with costs comparable to those of existing or near-term high-energy neutrino and gravitational-wave observatories.  As we discuss below, there are uncertainties in the supernova rates and neutrino emission.  It is expected that these uncertainties will be reduced by ongoing studies; in any case, direct new measurements of these quantities is precisely the goal of a detector as described here.

A principal goal of this paper is to open a discussion of supernova neutrino detection in very large detectors by presenting a reasonably detailed consideration of the science goals, detection aspects, and possible benefits of a detector large enough to routinely detect neutrino mini-bursts from supernovae in nearby galaxies.  Further work will be needed to develop the basic points of this paper.  The results and perspective for a $\sim\,$5~Mton detector are qualitatively different from previous work for even $\sim\,$1~Mton detectors, as in Refs.~\cite{Ando:2005ka,MTnu}, where typically one neutrino at a time is detected and a coincidence with an optical detection is required.  New possibilities emerge when neutrinos alone are sufficient to detect the core collapse and when the frequency of these detections is high.  Further, there are important questions about supernovae that can only be answered with many detected bursts, of which some can be answered with even a few detected neutrinos.

Before elaborating on details concerning detection rates, we will begin by exploring how the data obtained from multiple neutrino bursts would transform the way that we consider questions about supernovae; although this section is an overview, it contains several new points.  We will then examine recent developments concerning the rate and properties of supernovae observed in the nearby universe.  This will lead into our detailed discussion, much of it new, of the detector properties required to measure neutrino bursts from these supernovae and the quantitative new results on the mini-burst rates and neutrino yields expected.  While our treatment is based on the proposed parameters of Deep-TITAND~\cite{Suzuki:2001rb,Suzuki}, there could be other ways of constructing a multi-megaton detector for supernova neutrino bursts, and we encourage such studies.  An example is the consideration of a densely-instrumented infill array for the IceCube detector~\cite{ICtalk,ICtalk2}.  Finally, we present the overall conclusions and further discuss some specific highlights.


\section{Overview of Discovery Prospects}
\label{prospects}

Our primary interest is on the scientific impact of measuring neutrino ``mini-bursts,'' detectable signals of 3 or more events within 10 seconds (the observed duration of the SN~1987A neutrino burst), from many supernovae in the nearby universe.  As we will show in Sections~\ref{rate} and~\ref{detection}, the minimum detector size for achieving this purpose is about 5~Mton.  We emphasize in advance that such signals can be separated from backgrounds even at shallow depth, so that the presence of a core collapse can be deduced independently of photon-based observations.  Additionally, for nearby transients identified through photons, a non-detection in neutrinos means that a conventional supernova neutrino flux was not present.  These facts have new and profound implications.

While our principal focus is thus on individual objects, the aggregate data would, of course, also be useful.  For science goals that require a large number of accumulated events, the most certain signal is the Diffuse Supernova Neutrino Background (DSNB), which is a steady flux arising from all core-collapse supernovae in the universe (e.g., Refs.~\cite{DSNB,Yuksel:2007mn} and references therein).  In the proposed $\sim 0.5$ Mton HK detector, with added gadolinium to reduce backgrounds by neutron tagging~\cite{Beacom:2003nk}, $\sim\,$50--100 DSNB signal events with little background could be collected per year.  The ratio of DSNB signal to detector background in Deep-TITAND would be the same as in the background-dominated SK search of Ref.~\cite{Malek:2002ns}, which set an upper limit.  To reach the smallest plausible DSNB signals, one needs an improvement of about a factor 3 in signal sensitivity and thus a factor of about 10 in exposure.  After four years, as in the SK search, the Deep-TITAND exposure would be about 100 times larger than that of Ref.~\cite{Malek:2002ns}, thus allowing a robust detection of the DSNB flux.  (To measure the spectrum well, HK with gadolinium would be needed.)

\begin{table}[t!]
\caption{Approximate neutrino event yields for core-collapse supernovae from representative distances and galaxies, as seen in various detectors with assumed fiducial volumes.  Super-Kamiokande is operating, and Hyper-Kamiokande and Deep-TITAND are proposed.}
\label{tab:detectors}
\begin{ruledtabular}
\begin{tabular}{llccc}
	&	& 32 kton	& 0.5 Mton	& 5 Mton \\
	&	& (SK)	& (HK)		& (Deep-TITAND) \\ \hline
10 kpc   & (Milky Way)  & $10^4$    & $10^5$   & $10^6$ \\
1 Mpc    & (M31, M33)  & $1$       & $10$     & $10^2$ \\
3 Mpc    & (M81, M82)  & $10^{-1}$ & $1$      & $10$ \\
\end{tabular}
\end{ruledtabular}
\end{table}

The fortuitous occurrence of a supernova in the Milky Way would obviously result in an abundance of neutrino events (see Table~\ref{tab:detectors}) and the physics prospects associated with such yields from a single supernova have been discussed for underground detectors at the 0.5~Mton scale~\cite{MTnu}.  However, even Andromeda (M31) or Triangulum (M33) would give $\sim 100$ neutrino events.  The physics prospects associated with yields of $\sim\,$10 events for these galaxies, comparable to SN~1987A, have been discussed for $\sim\,$0.5~Mton underground detectors.  With $\sim\,$10 times more events, a substantial improvement over the results of SN~1987A should be possible.  Further, bursts comparable to SN~1987A would be more common.  For example, M82, a nearby starburst galaxy, is thought to have a supernova rate as large as 10 per century~\cite{Rieke}, and there are other galaxies within its distance range.


\subsection{Probing the core collapse mechanism}

The optical signals of supposed core-collapse supernovae show great diversity~\cite{Zwicky(1940), Filippenko:1997ub}, presumably reflecting the wide range of masses and other properties of the massive progenitor stars.  In contrast, the neutrino signals, which depend on the formation of a $\sim 1.4\, M_\odot$ neutron star, are presumed to be much more uniform.  However, since we have observed neutrinos only from SN~1987A, it remains to be tested whether all core-collapse supernovae do indeed have comparable neutrino emission.  The total energy emitted in neutrinos is $\simeq 3 G M^2 / 5 R$, and some variation is expected in the mass $M$ and radius $R$ of the neutron star that is formed, though proportionally much less than in the progenitor stars.

With at least $\sim 1$ nearby supernova per year, a wide variety of supernovae can be probed, including less common types.  For example, the observational Types Ib and Ic are now believed to be powered by core collapse, despite their original spectroscopic classification that defined them as related to Type~Ia supernovae, which are thought to be powered by a thermonuclear runaway without significant neutrino emission.  While each of the Types Ib/Ic and Ia are only several times less frequent than Type~II, some of each should occur nearby within a reasonable time, so that the commonality of the Type II/Ib/Ic explosion mechanism can be tested.

While the nature of the explosion in the above supernova types is very likely as expected, there are other bright transients observed for which the basic mechanism is much more controversial.  For these events, we make the new point that the detection or non-detection of neutrinos could decisively settle debates that are hard to resolve with only optical data.  One type of so-called ``supernova impostor'' is thought to be the outburst of a Luminous Blue Variable (LBV)~\cite{Humphreys}, which seems to require a stellar mass of $M_* \gtrsim 20\,M_\odot$.  Since this type of outburst affects only the outer layers, with the star remaining afterward, there should be no detectable neutrino emission.

There are several recent examples in nearby galaxies where neutrino observations could have been conclusive, including the likely LBV outburst SN~2002kg in NGC~2403~\cite{VanDyk:2006ay}.  SN~2008S in NGC~6946~\cite{Prieto:2008bw} and a mysterious optical transient in NGC~300~\cite{Thompson08} warrant further discussion for another reason.  In neither case was a progenitor seen in deep, pre-explosion optical images; however, both were revealed as relatively low-mass stars ($M_* \sim 10\,M_\odot$) by mid-infrared observations made years before the explosions.  This suggests that they were obscured by dust expelled from their envelopes, a possible signature of stars dying with cores composed of O-Ne-Mg instead of iron~\cite{Prieto:2008bw,Thompson08}, which may lead to unusual neutrino mixing effects~\cite{Lunardini:2007vn}.  As we will address in detail later, these events were sufficiently near for a 5~Mton detector to have identified them as authentic supernovae or impostors.


\subsection{Measuring the total core collapse rate}

In the previous subsection, we implicitly considered supernovae for which the optical display was seen.  However, as we will calculate, the detection of $\ge 3$ neutrinos is sufficient to establish that a core collapse occurred, including those events not later visible to telescopes.  This provides a means of measuring the total rate of true core collapses in the nearby universe.

A successful supernova may be invisible simply if it is in a very dusty galaxy, of which there are examples quite nearby, such as NGC~253 and M82.  These are supposed to have very high supernova rates, perhaps as frequent as one per decade each, as deduced from radio observations of the number of young supernova remnants~\cite{Muxlow}.  However, only a very few supernovae have been seen~\cite{CBAT}.  A recent example is SN~2008iz, which was not seen in the optical, being detected only via serendipitous radio observations~\cite{Brunthaler:2010bm}.  This is exactly the kind of event for which a neutrino detector would be especially useful, as it can monitor all directions at once to find core collapses that would otherwise be missed.

More interestingly, it remains unknown if, as in numerical models of supernova explosions, some core collapses are simply not successful at producing optical supernovae.  This can occur if the outgoing shock is not sufficiently energetic to eject the envelope of the progenitor star, in which case one expects the prompt formation of a black hole with very little optical emission~\cite{Heger:2002by}.  Indirect evidence for such events follows from a deficit of high-mass supernova progenitors compared to expectations from theory~\cite{Kochanek:2008mp,Smartt:2008zd}, as well as from the existence of black holes recently discovered to have $M_{\rm BH}\gtrsim 15\,M_\odot$~\cite{Orosz:2007ng}.

One way to probe this exotic outcome would be to simply watch the star disappear as an ``unnova''~\cite{Kochanek:2008mp}.  However, a detectable burst of neutrinos should be emitted before the black hole forms (and typically, if the duration of the emission is shorter, the luminosity is higher)~\cite{Burrows:1988ba,Beacom:2000qy,Sumiyoshi:2008zw,Keehn:2010pn}.  Taken together, these would be a dramatic and irrefutable signal of an otherwise invisible event, and it is a new point that the detection of bursts from core collapses in nearby galaxies could be a practical way to probe even small rates of black hole forming collapses and their resulting neutrino spectra.  While the rate of prompt black hole formation probably cannot exceed the visible supernova rate without violating constraints on the DSNB, reasonable estimates indicate that up to $\gtrsim 20\%$ of core collapses may have this fate~\cite{Kochanek:2008mp}.


\subsection{Testing the neutrino signal}

By measuring neutrinos from many supernovae, the deduced energy spectra and time profiles could be compared to each other and to theory.  In most cases, only several events would be detected, but this is enough to be useful.  The highest neutrino energies range up to $\simeq 50$ MeV.  The thermal nature of the neutrino spectrum makes it relatively narrow, and since it is falling exponentially at high energies, even a small number of events can help determine the temperature.  Recall that for SN~1987A, the Kamiokande-II and IMB detectors collected only $\sim 10$ events each~\cite{Hirata:1987hu, Bionta:1987qt}, but that this data strongly restricts the details of the collapsed core.  

The time profile is thought to rise quickly, over perhaps at most 0.1 s, and then decline over several seconds, as seen for SN~1987A.  The neutrino events collected would most likely be at the early peak of the emission, and hence the most relevant for the question of whether heating by the emergent neutrino flux is adequate for shock revival~\cite{Bethe:1984ux,Thompson:2002mw,Murphy:2008dw} or whether $\nu$-$\nu$ many-body effects are important~\cite{nu-nu}.

Over time, as many supernovae are detected, the average energy spectrum and time profile will be built up.  (For the time profile, there will be some uncertainties in the start times.)  If there are large variations from one supernova to the next, then these average quantities will ultimately provide a more useful template for comparison than the theoretical results that must be used at present.  If there is no evidence for significant variations between supernovae, then the accumulated data will be equivalent to having detected one supernova with many events.  It is quite likely that such a detector would observe a supernova in one of the Milky Way, M31, or M33; the high-statistics yield from these would also provide a point of comparison.  Taken together, all of these data will provide new and exacting tests of how supernovae work.  We note that it is hard to imagine any other way to test the variation in neutrino emission per supernova.

With enough accumulated events, it is expected that neutrino reactions besides the dominant inverse beta decay process will be present in the data.  One oddity still remaining from SN~1987A is that the first event in Kamiokande-II seems to be due to $\nu_e + e^-\rightarrow \nu_e + e^-$ scattering and points back to the supernova~\cite{Hirata:1988ad}, which is improbable based upon standard expectations~\cite{Beacom:2006}.  This can be tested, however, and if it turns about to be ubiquitous, could be exploited in determining the directionality of the larger future bursts without optical signals, as the inverse beta decay signal is not directional~\cite{Vogel:1999zy}.

Since Earth is transparent to supernova neutrinos, the whole sky can be monitored at once.  For neutrinos that pass through Earth, particularly those which cross the core, matter-enhanced neutrino mixing can significantly affect the spectrum relative to those which do not~\cite{earth}.  Dividing the accumulated spectra appropriately based on optical detections, this would allow a new test of neutrino mixing, sensitive to the sense of the neutrino mass hierarchy~\cite{Dighe:1999bi}.  Detecting neutrinos from distant sources would also allow tests of neutrino decay~\cite{decay}, the equivalence principle~\cite{equivalence}, and other exotic possibilities~\cite{exotic}.


\subsection{Revealing other transient signals}

Detection of a neutrino burst means detection of the instant of core collapse, with a precision of $\sim 1$ second determined by the sampling of the peak of the $\simeq 10$ second time profile.  This would provide a much smaller time window in which to search for gravitational wave signals~\cite{Ando:2005ka,LIGO,GW,Halzen:2009sm} from core-collapse supernovae; otherwise, one must rely on the optical signal of the supernova, which might optimistically be determined to a day ($\sim 10^5$ seconds).  This is important, since the gravitational-wave signal remains quite uncertain, making searches more difficult.  Knowing the instant of core collapse would also be useful for searches for high-energy neutrinos from possible choked jets that do not reach the surface of the star~\cite{Jets}, where again the timing information can be used to reduce backgrounds and improve sensitivity.

Once core collapse occurs, the outward appearance of the star initially remains unchanged.  Knowing that a signal was imminent would give advanced warning, as previously discussed for a Galactic supernova (e.g., Ref.~\cite{Antonioli:2004zb}), that photons should soon be on the way.  This allows for searches to commence for the elusive UV/X-ray signal of supernova shock breakout~\cite{SBO} and also the early supernova light curve.  Those signals are expected to emerge within a period of minutes to days, depending upon the progenitor star.  While the neutrino signal will likely not provide directional information, the number of events detected will provide constraints for triggered searches, providing a new way to improve the chances of early electromagnetic detection of extragalactic supernovae~\cite{Kistler}.

Finally, it is possible that a large detector would find not only core-collapse supernovae in nearby galaxies, but also other types of neutrino transients that are presently unknown.  Mergers involving compact objects could lie in this class~\cite{Popham:1998ab,Rosswog:2003rv,Kalogera:2003tn}.  In the Milky Way, there would be sensitivity to any transient with a supernova-like neutrino signal, as long as its overall strength is at least $\sim 10^{-6}$ as large as that for a supernova.  To be detectable, the key requirement is a $\gtrsim 15$~MeV $\bar{\nu}_e$ component.


\section{Nearby Supernova Rate}
\label{rate}

Over the past decade, there has been rapid growth in the level of interest among astronomers in measuring the properties of core-collapse supernovae.  There is also a renewed interest in completely characterizing the galaxies in the nearby universe, within 10 Mpc.  In nearby galaxies, both amateurs and automated surveys (e.g., KAIT/LOSS~\cite{Li:1999sd,Leaman:2010kb}) are finding many supernovae.  For these SNe, archival searches have revealed pre-explosion images of about a dozen supernova progenitor stars, allowing a better understanding of which types of massive stars lead to which kinds of core-collapse supernovae (e.g.,~\cite{Prieto:2008bw,Smartt:2008zd,Li07,GalYam:2006iy}).

\begin{figure}[t!]
\includegraphics[width=3.25in,clip=true]{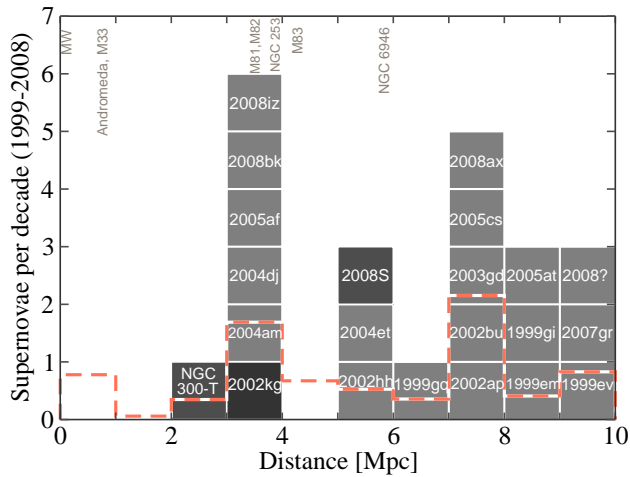}
\caption{Estimates of the core-collapse supernova rate in the nearby universe, based on that expected from the optical luminosities of known galaxies (line) and 22 supernovae observed in 1999-2008 (bins).  Note that SN 2002kg is a likely LBV outburst, while SN 2008S and the NGC 300 transient are of unusual origin.  These estimates are all likely to be incomplete.}
\label{fig:snrates}
\end{figure}

Figure~\ref{fig:snrates} shows the expected rate of core-collapse supernovae in the nearby universe (dashed line) calculated using the galaxy catalog of Ref.~\cite{Karachentsev:2004dx} (designed to be $\sim$70--80\% complete up to 8~Mpc), with a conversion from measured $B$-band optical luminosity to supernova rate from Ref.~\cite{Cappellaro:1999qy}.  Within 10~Mpc, there are $\sim\,$40 major galaxies that contribute most of the expected supernova rate; the most important ones are listed in Ref.~\cite{Kochanek:2008mp}; we include the many dwarf galaxies in the catalog, though this has only a modest effect on the total rate.  The effects of clustering and of incompleteness at large distances can clearly be seen, since the histogram would rise as the distance squared for a smooth universe of identical galaxies.  The conversion from measured galaxy luminosity in the B-band to estimated supernova rate involves multiplication by an empirical conversion factor (see Refs.~\cite{Cappellaro:1999qy,Ando:2005ka,Horiuchi:2011zz} for details on the uncertainty).  The essential problems with using the B-band light as a measure of high-mass stars and hence the core-collapse supernova rate are the variations in the correlation due to dust obscuration and the mix of high-mass and low-mass stars.  Ultimately, a more accurate result could be obtained by combining the information from star-formation rate measurements in the ultraviolet~\cite{Salim:2004dg}, H$\alpha$~\cite{Halpha}, and infrared~\cite{Kennicutt:2003dc}, likely leading to a larger prediction for the supernova rates.

We can avoid the above uncertainties by directly using measured supernova rates in nearby galaxies, which gives an example of what nature has provided in the past.  Displayed in Fig.~\ref{fig:snrates} is the rate deduced from supernovae discovered in this volume in 1999-2008~\cite{CBAT}, with distances primarily from Ref.~\cite{Karachentsev:2004dx} (when available; otherwise from~\cite{WEBdist}).  While the observed rate is already $\sim 2$ times larger than the above calculation, even this estimate is likely incomplete, as supernova surveys under-sample small galaxies and the Southern hemisphere.  The recent archival discovery of a bright Type~II SN in a $\sim\,$9.5~Mpc galaxy missed by targeted surveys (denoted as SN~2008?~in Fig.~\ref{fig:snrates}) provides direct evidence in this direction~\cite{Prieto11}.  As previously mentioned, supernovae with little or no optical signal, e.g., due to direct black hole formation or dust obscuration, would also have been missed~\cite{Ando:2005ka,Brunthaler:2010bm,Kochanek:2008mp,Horiuchi:2011zz}.  This is particularly important for nearby dusty starburst galaxies with large expected, but low observed, supernova rates, like NGC~253 and M82.

Distance measurements of nearby galaxies also stand to be improved.  For example, at the largest distances, SN~1999em, SN~1999ev, SN~2002bu, and SN~2007gr may not all truly reside within 10~Mpc, as some distance measures place them outside.  We emphasize that their inclusion or not does not affect our approximate supernova rates, and barely matters for the neutrino bursts of sufficient multiplicity, which are dominantly from closer supernovae. It would be very helpful to refine distance measurements, not just for star formation/supernova rate estimates, but also to determine the absolute neutrino luminosities once a supernova has been detected.

\begin{figure}[b]
\includegraphics[width=3.25in,clip=true]{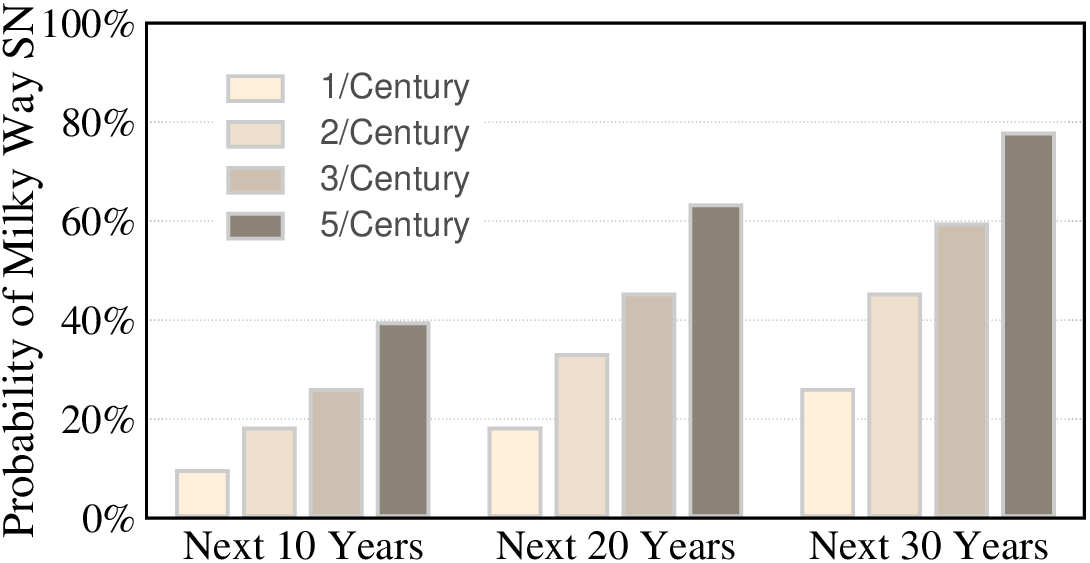}
\caption{Probabilities for one or more supernovae in the Milky Way over time spans relevant for the lifetimes of large neutrino detectors, depending on the assumed supernova rate.}
\label{fig:forecast}
\end{figure}

Overall, there is a good case that the core-collapse rate within $\sim $ 6 (10) Mpc is at least 1 (2) per year.  We expect that ongoing studies of star formation and supernova rates in nearby galaxies can reduce the uncertainty.  However, even for a known average rate, there will remain relatively large Poisson uncertainties on the actual rate during short periods even in the whole collection of nearby galaxies, which limits the level of refinement in the predictions.  This rate can be compared to the estimated Milky Way rate of $2 \pm 1$ per century (see Ref.~\cite{Diehl:2006cf} and references therein), with Poisson probabilities ultimately determining the odds of occurrence, as shown in Fig.~\ref{fig:forecast}.

\section{Neutrino Burst Detection}
\label{detection}

A goal of measuring supernova neutrino ``mini-bursts'' from galaxies at a few Mpc necessitates a large detector, roughly $\sim$100 times the size of SK.  We focus on the Deep-TITAND proposal for a 5 Mton (fiducial volume) enclosed water-\v{C}erenkov detector~\cite{Suzuki:2001rb,Suzuki}.  The detector would be constructed in modules sized by \v{C}erenkov light transparency and engineering requirements.  We assume a photomultiplier coverage of $20\%$, similar to that of SK-II (half that of the original SK-I and the rebuilt SK-III).  As in SK, the detection efficiency at the energies considered here would be nearly unity.

The backgrounds present in deep detectors have been well-characterized by SK and other experiments.  Deep-TITAND is proposed to be at a relatively shallow depth of 1000 meters of seawater, which would increase the downgoing cosmic ray muon rate per unit area by a factor $\simeq 100$ compared to SK, which is at a depth of 2700 meters water equivalent.  A nearly perfect efficiency for identifying cosmic ray muons in the outer veto or the detector itself is required.  This was achieved in SK, where the only untagged muons decaying in the detector were those produced inside by atmospheric neutrinos~\cite{Malek:2002ns}.  Simple cylinder cuts around cosmic ray muon tracks would veto all subsequent muon decays while introducing only a negligible detector deadtime fraction.

Low-energy backgrounds include natural radioactivities, solar neutrinos, photomultiplier noise, and beta decays from nuclei produced following spallation by cosmic ray muons.  Of these, only the last is depth-dependent, and this would be much larger than in SK (a factor $\simeq 30$ for the higher muon rate per area but lower muon average energy, and a factor $\simeq 30$ for the larger detector area).  The high muon rate means that it would not be possible to use the same cylinder cuts employed in SK to reduce spallation beta decays without saturating the deadtime fraction (note that these beta decays have lifetimes more than $10^6$ times longer than the muon lifetime).  At low energies, the above background rates are large, but the spectrum falls steeply with increasing energy, essentially truncating near 18 MeV~\cite{Malek:2002ns, Ikeda:2007sa}.

This allows for a significant simplification and reduction in the background rate by considering only events with a reconstructed energy greater than 18 MeV (a neutrino energy of 19.3 MeV).  Which events to reconstruct would be determined by a simple cut on the number of hit photomultipliers, just as in SK, but with a higher threshold.  The backgrounds above this cut are due to atmospheric neutrinos, and thus the rates scale with the detector volume but are independent of depth.  The dominant background contribution is from the decays of non-relativistic muons produced by atmospheric neutrinos in the detector, i.e., the so-called invisible muons.  The background rate in 18--60 MeV in SK is about 0.2 events/day, of which the energy-resolution smeared tail of the low-energy background is only a minor component~\cite{Malek:2002ns, Ikeda:2007sa}.

Scaling this rate to a 5 Mton detector mass ($\sim 5 \times 10^{-4}$ s$^{-1}$) and considering an analysis window of 10~sec duration (comparable to the SN~1987A neutrino signal) allows calculation of the rate of accidental coincidences~\cite{Ikeda:2007sa}.  For $N = 3$ events, this corresponds to about only once every five years, and when it does, examination of the energy and timing of the events will allow further discrimination between signal and background (a subsequent optical supernova would confirm a signal, of course).  For $N \ge 4$, accidental coincidences are exceedingly rare ($\sim\,$1 per 3000~years), therefore we require at least $N = 3$ signal events to claim detection of a supernova (a somewhat greater requirement than in Ref.~\cite{Ando:2005ka}, where a smaller detector was assumed).  Since the backgrounds observed by SK in this energy range are from atmospheric neutrinos, we expect no correlated clusters of background events.

To estimate detection prospects, for the $\bar{\nu}_e$ flavor we assume a Fermi-Dirac spectrum with an average energy of 15 MeV and a total energy of $5\times 10^{52}$~erg.  These are reasonable values for the effective received spectrum of $\bar{\nu}_e$ after neutrino mixing in the supernova.  In many theoretical papers, significantly larger values for these parameters after neutrino mixing are assumed.  We can also make a comparison to the SN~1987A data.

As our calculations below depend on the positron spectrum above 18 MeV, only the higher-energy SN~1987A data, primarily the events seen in the IMB detector,  are relevant for estimating the received spectrum.   Thermal fits to the shape of the high-energy spectrum~\cite{Jegerlehner:1996kx,Mirizzi:2005tg,Lunardini:2004bj,Pagliaroli:2008ur} or direct reconstruction thereof~\cite{Yuksel:2007mn} are in reasonable agreement with this assumed spectrum.  The thermal fits allow lower average energies if accompanied by higher total energies.  Those fits, just like the predictions below, depend on the number of detected events, which is approximately the product of the average and total energies.  Of course, we do not know if SN 1987A was typical, and testing this is one of the goals of such a large detector.

\begin{table}[t!]
\caption{Core-collapse supernova candidates from 1999-2008 within 6~Mpc, with their expected neutrino event yields ($E_{e^+} > 18$~MeV) in a 5 Mton detector.}
\label{tab:yields}
\begin{ruledtabular}
\begin{tabular}{lcccr}
SN     & Type &  Host     & D [Mpc] & $\nu$ events\\ \hline
2002hh & II-P &  NGC 6946 & 5.6     & 2.4\\
2002kg & IIn/LBV & NGC 2403 & 3.3   & 6.8\\
2004am & II-P &  NGC 3034 (M82) & 3.53   & 5.9\\
2004dj & II-P &  NGC 2403 & 3.3     & 6.8\\
2004et & II-P &  NGC 6946 & 5.6     & 2.4\\
2005af & II-P &  NGC 4945 & 3.6     & 5.7\\
2008S  & IIn  &  NGC 6946 & 5.6     & 2.4\\
2008bk & II-P &  NGC 7793 & 3.91    & 4.8\\
2008iz & II? &  NGC 3034 (M82) & 3.53   & 5.9\\
NGC 300-T & II?  &  NGC 300  & 2.15    & 16.0\\
\end{tabular}
\end{ruledtabular}
\end{table}

The dominant interaction for the neutrino signal is inverse-beta decay, $\bar{\nu}_e + p\rightarrow n + e^+$, where $E_{e^+} \simeq E_{\bar{\nu}_e} - 1.3$ MeV and the positron direction is nearly isotropic~\cite{Vogel:1999zy}.  Combining the emission spectrum, cross section, and number of free target protons in a water detector of mass $M_{\rm det} = 5$~Mton, we find that the average number of neutrino events (for $E_{e^+} > 18$ MeV) from a burst at distance $D$ is
\begin{equation}
\mu (D; E_{e^+} > 18~\mathrm{MeV}) \simeq 5\, \left(\frac{M_{\rm det}}{5\, \mathrm{Mton}}\right) \left(\frac{3.9\, \mathrm{Mpc}}{D}\right)^{2}.
\end{equation}
This is the key normalization for the supernova signal.  In Table~\ref{tab:yields}, we list recent nearby supernovae within 6~Mpc, with type, host galaxy name, distance, and the expected neutrino yields $\mu$ in a 5~Mton detector.  As can be seen in Fig.~3 of Ref.~\cite{Ando:2005ka}, our $E_{e^+} > 18$ MeV threshold still allows us to detect $\sim 70\%$ of the total supernova signal.

The probability to detect $\geq N$ neutrino events from a given core collapse is then
\begin{equation}
P( \geq N; D) = \sum_{n = N}^{\infty} P_n [\mu (D)] = \sum_{n=N}^\infty \frac{\mu^n(D)}{n!}e^{-\mu (D)},
\end{equation}
where $P_n(\mu)$ represents the Poisson probability.  $P(\geq N; D)$ is shown in Fig.~\ref{fig:yields} as a function of $D$ for several values of $N$.  From this figure, we see, for example, that from a 4~Mpc supernova, we have an excellent chance ($\gtrsim 90\%$) to get more than 3 neutrino events.  For 8~Mpc, like those shown in Fig.~\ref{fig:snrates}, there is still a $\lesssim 10\%$ chance to get $\ge 3$ events.

\begin{figure}[b!]
\includegraphics[width=3.25in,clip=true]{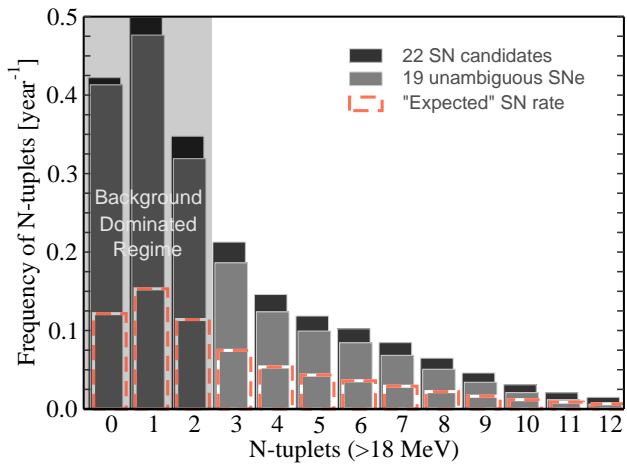}
\caption{Frequency of neutrino mini-bursts expected with a 5~Mton detector.  The bins with $N = 3$ or more can be used for burst detection because the background rate is small enough.  Three different estimates of the supernova rate are shown, as labeled.}
\label{fig:multiplicities}
\end{figure}

For a particular supernova rate, $R_{{\rm SN},i}$, we can get the expected total rate of $N$-tuplet detections from distances $D_i$ as
\begin{equation}
R_{N,{\rm burst}} = \sum_i R_{{\rm SN},i} P_N [\mu (D_i)]\,,
\end{equation}
where the sum runs over the list of nearby galaxies.   This sum form is more accurate than an integral form that forces a continuum limit.  In Fig.~\ref{fig:multiplicities}, we show this as a yearly rate, $R_{N,{\rm burst}}$, plotted versus $N$.  For the supernova rate $R_{{\rm SN},i}$, we have adopted three different models: (i) all supernova candidates shown in Fig.~\ref{fig:snrates} (22 in total); (ii) same as (i), except excluding SN~2002kg, SN~2008S, and the NGC~300 transient as exceptional events (19 in total); (iii) a catalog-based rate estimate corresponding to the line in Fig.~\ref{fig:snrates} for galaxies at $D>2$~Mpc.  In the first two cases, the rates depend on integer numbers of observed supernovae; in the third, the rate depends on the ``expected'' (fractional) number of supernovae.  As the detection criterion is $N \geq 3$, the rate of detectable mini-bursts is obtained by summing $R_{N,{\rm burst}}$ for $N \ge 3$, which yields 0.9, 0.7, and 0.3 mini-bursts per year, for supernova rate models (i), (ii), and (iii), respectively.  Supernovae from beyond 10~Mpc do not appreciably change the rate of $N \ge 3$ multiplets, only increasing the number of unremarkable lower-$N$ multiplets (which, as shown, are already dominated by supernovae in the 8--10 Mpc range) and can be regarded as a component of the DSNB.  We emphasize that we view the case (iii) as too conservative, as it significantly underpredicts the number of core collapse events, likely by a factor of $\sim\,$2 (for a fuller discussion, see Ref. \cite{Horiuchi:2011zz}), and also note that none of (i), (ii), and (iii) can account for failed supernovae.

The total neutrino event counts, $N_{\rm total}$, from mini-bursts with $N \geq 3$ events is obtained from $R_{N,{\rm burst}}$ by
\begin{equation}
 N_{\rm total} = \sum_{N = 3}^{\infty} N R_{N,{\rm burst}}\,,
\end{equation}
which are 62, 37, and 22 per decade, for rate estimates (i), (ii), and (iii), respectively.  Since each burst is triggered with $E_{e^+} > 18$~MeV events, one would also look for somewhat lower-energy events in the same time window, potentially raising the total yield by $\simeq 20\%$.

As noted above, (iii) does not include galaxies at distances $< 2$~Mpc, as we have focused on the frequency of detectable mini-bursts.  A detector of this type would surely run for at least a few decades, long enough to make it quite probable that a supernova occurs in one of the Milky Way, M31, M33, or their smaller satellite galaxies; see Table~\ref{tab:detectors} for approximate distances and yields.  Importantly, this would mean that at least one burst would be detected with $\gtrsim\,$100 events and possibly much more, significantly increasing the scientific return.  The high signal rates would mean that events below 18~MeV could be used, raising the overall yields, giving a better measure of the spectrum, and possibly including events besides those from the inverse-beta detection channel.

\begin{figure}[t!]
\includegraphics[width=3.25in,clip=true]{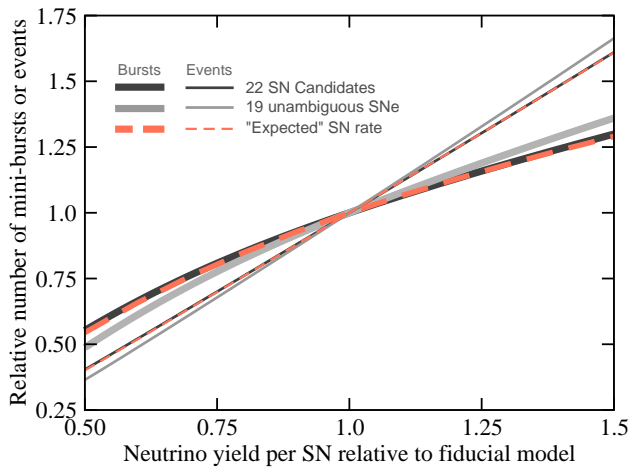}
\caption{Relative number of $N \geq 3$ neutrino mini-burst detections and summed neutrino counts as the expected neutrino event yield in Eq.~(1) is varied from our fiducial case of a Fermi-Dirac $\bar{\nu}_e$ spectrum with an average energy of 15 MeV and a total energy of $5\times 10^{52}$~erg with a 5~Mton water detector.  A range of from $0.7-1.3$ on the horizontal axis can be roughly estimated from the high-energy neutrino events observed from SN~1987A (see text).}
\label{fig:bok}
\end{figure}

To understand the uncertainty on the overall normalization that we have used in Eq.~(1), note that only the number of events above a positron energy of 18 MeV is needed, independent of the shape of the spectrum in this range.  We calculated how the normalization in Eq.~(1) depends on variations about our assumed parameters for the received $\bar{\nu}_e$ emission spectrum.  Varying the total energy alone leads to a relative change in the normalization of the same size.  Changing the average energy alone leads to a relative change in the normalization that is nearly linear but about twice as large.  As noted above, our assumed normalization depends on both parameters in the same way as the number of high-energy events from SN~1987A.  As variations in the two parameters can have compensating effects, we consider a combined uncertainty on the normalization of Eq.~(1).  A rough Poisson uncertainty of $\sim\,$30\% can be deduced from the $\sim\,$10 high-energy events from SN 1987A, although it is difficult to assess an uncertainty on the typicality of SN~1987A.

As shown in Fig.~\ref{fig:bok}, we find by direct calculation that the changes in our results are nearly linear with variations in the uncertainty on the normalization of Eq.~(1), which makes it easy to estimate the effects of alternate assumptions concerning the supernova neutrino emission (e.g., average energy, luminosity, oscillatory effects, etc.).  As we have only included supernovae within 10~Mpc, the curves displayed should be considered underestimates for event yields larger than our fiducial case (the region $>1$ on the horizontal axis); the size of this possible underestimate can be gauged from Fig.~\ref{fig:yields}.  This could arise from the core-collapse neutrino emission being larger than assumed here.  The uncertainties on our results are not unduly magnified from the uncertainties on supernova emission parameters by the exponentials in the thermal spectrum and Poisson probability because the energy and count cuts are comparable to the average expectations.  Since Fig.~\ref{fig:bok} shows how the number of bursts/events varies when the neutrino yield changes by up to 50\% with respect to the fiducial model regardless of the source of the change, this also allows for a more general examination for water detector masses other than 5~Mton, of particular relevance for scalable detector designs.


\section{Discussion and Conclusions}

The $\sim 10$ neutrino events associated with SN 1987A in each of the Kamiokande-II and IMB detectors~\cite{Hirata:1987hu,Bionta:1987qt} were the first and, thus far, only detection of neutrinos from a supernova.  This detection showed that we can learn a great deal even from a small number of events, and revealed that an immense amount of energy is released in the form of neutrinos ($> 10^{53}$~erg) during a core collapse.  Measuring ``mini-bursts'' of neutrino events from multiple supernovae would allow for the study of the core-collapse mechanism of a diverse range of stellar deaths, including optically-dark bursts that appear to be relatively common~\cite{Kochanek:2008mp, Smartt:2008zd}.

This would be made possible by a $\sim\,$5~Mton scale water \v{C}erenkov detector~\cite{Suzuki:2001rb, Suzuki}, which has the special advantages of being able to trigger on supernovae using neutrinos alone, and to guarantee detection if neutrinos are produced with the expected flux.  Moreover, for burst detection, a relatively-high low-energy background rate can be tolerated, significantly decreasing the required detector depth, so that construction could be relatively quick and inexpensive.  Such direct measurements with neutrinos will ultimately be needed to resolve the important questions discussed in Section II.

Our estimates show that the occurrence rate of mini-bursts that give $\ge 3$ neutrino events is at least several per decade.  Because neutrinos will be detected in bursts, it will be possible to separately explore questions about the neutrino emission per core collapse and the core collapse rate.  A detector of this type would run for decades, and would accumulate neutrino statistics with at least the yearly rates mentioned in the previous section.  There would also be a good chance of seeing a large burst from a supernova in M31 or M33 ($\sim\,10^2$ events), the  Milky Way ($\sim\,10^6$ events), or their satellite galaxies.

In conclusion, we wish to reiterate that, even if a supernova occurs in the Milky Way tomorrow, the important problems discussed in Section~\ref{prospects} will remain unresolved, and can only be addressed with certainty by a suitable ``census'' of core collapses in the nearby universe.  The possibilities mentioned here almost certainly do not exhaust the scientific potential of such an instrument.  As is now almost commonplace in the business of observing supernovae with photons, it would be surprising {\it not} to find new and unexpected phenomena.


{\bf Acknowledgments:}
We thank Shunsaku Horiuchi, Chris Kochanek, Y. Ohbayashi, Jos\'{e} Prieto, Stephen Smartt,
Michael Smy, Kris Stanek, Todd Thompson, and Mark Vagins for helpful discussions.
This work was supported by Department of Energy grant DE-FG02-91ER40690
(MDK); National Science Foundation CAREER grant PHY-0547102 to JFB (HY
and JFB); and by the Sherman Fairchild Foundation at Caltech (SA).


\end{document}